\documentclass[superscriptaddress, amsmath, aps, prl, numerical, twocolumn, nofootinbib]{revtex4-1}

\usepackage{graphicx}
\usepackage{dcolumn}
\usepackage{bm}
\usepackage{xcolor}

\newcommand{\comment}[1]{}

\begin{document}

\title{Narrow resonances in the continuum of the unbound nucleus $^{15}$F}

\author{V. Girard-Alcindor}
\affiliation{  Grand Accélérateur National d’Ions Lourds (GANIL), CEA/DRF-CNRS/IN2P3, Bvd Henri Becquerel, 14076, Caen, France}
\affiliation{    IJCLab, Université Paris-Saclay, CNRS/IN2P3, 91405, Orsay, France}
\author{A. Mercenne}
\affiliation{Center for Theoretical Physics, Sloane Physics Laboratory, Yale University, New Haven, Connecticut 06520, USA}
\author{  I. Stefan}
\affiliation{    IJCLab, Université Paris-Saclay, CNRS/IN2P3, 91405, Orsay, France}
\author{   F. de Oliveira Santos}
\affiliation{  Grand Accélérateur National d’Ions Lourds (GANIL), CEA/DRF-CNRS/IN2P3, Bvd Henri Becquerel, 14076, Caen, France}
\author{   N. Michel}
\affiliation{ Institute of Modern Physics, Chinese Academy of Sciences, Lanzhou 730000, China}
\author{ M. P{\l}oszajczak}
\affiliation{  Grand Accélérateur National d’Ions Lourds (GANIL), CEA/DRF-CNRS/IN2P3, Bvd Henri Becquerel, 14076, Caen, France}
\author{ M.	Assié}
\affiliation{    IJCLab, Université Paris-Saclay, CNRS/IN2P3, 91405, Orsay, France}
\author{  A.	Lemasson}
\affiliation{  Grand Accélérateur National d’Ions Lourds (GANIL), CEA/DRF-CNRS/IN2P3, Bvd Henri Becquerel, 14076, Caen, France}
\author{  E.	Clément}
\affiliation{  Grand Accélérateur National d’Ions Lourds (GANIL), CEA/DRF-CNRS/IN2P3, Bvd Henri Becquerel, 14076, Caen, France}
\author{  F.	Flavigny}
\affiliation{ Normandie Univ, ENSICAEN, UNICAEN, CNRS/IN2P3, LPC Caen, 14000 Caen, France}
% ...

\date{\today}

\begin{abstract}
  \noindent
  The structure of the unbound $^{15}$F nucleus is investigated using the inverse kinematics resonant scattering of a radioactive $^{14}$O beam impinging on a CH$_2$ target.
  The analysis of $^{1}$H($^{14}$O,p)$^{14}$O and $^{1}$H($^{14}$O,2p)$^{13}$N reactions allowed the confirmation of the previously observed narrow $1/2^{-}$ resonance, near the two-proton decay threshold, and the identification of two new narrow 5/2$^{-}$ and 3/2$^{-}$ resonances. The newly observed levels decay by 1p emission to the ground of $^{14}$O, and by sequential 2p emission to the ground state (g.s.) of $^{13}$N via the $1^-$ resonance of $^{14}$O. Gamow shell model (GSM) analysis of the experimental data suggests that the wave functions of the 5/2$^{-}$ and 3/2$^{-}$ resonances may be collectivized by the continuum coupling to nearby 2p- and 1p- decay channels. The observed excitation function $^{1}$H($^{14}$O,p)$^{14}$O and resonance spectrum in $^{15}$F are well reproduced in the unified framework of the GSM.
  \end{abstract}

  \pacs{Valid PACS appear here}

  \maketitle

  \textit{Introduction.--} The nucleus is an open quantum system (OQS) where virtual excitations to continuum states provide an essential mechanism of the effective interaction \cite{GSMbook,Johnson2020}. 
   Well known manifestations of nuclear openness are segregation of decay time scales \cite{segr1,segr2}, 
  modification of the effective interactions \cite{GSMbook}, multichannel effects in reaction cross-sections and shell occupancies \cite{hate78,PhysRevC.75.031301}, or near-threshold clustering and correlations \cite{okolowicz2012,okolowicz2013}, etc. The latter phenomenon is generic in OQSs 
  and stems from properties of the scattering matrix in a multichannel system \cite{Zeldovich}. The coupling of different shell model (SM) eigenstates with the same quantum numbers (angular momentum and parity) to the same decay channel induces a mixing among them, reflecting the nature of the decay channel \cite{okolowicz2012,okolowicz2013}. Such configuration mixing  can radically change the structure of near-threshold states.

  Resonance spectroscopy of nuclei located far from the valley of stability and close, or beyond, the neutron and proton driplines, is the basic experimental tool to study coupling of discrete states with a scattering continuum. Due to their extremely short half-life, these nuclei are difficult to produce and study. Investigating the structure of the excited states of these unbound nuclei is an even greater challenge, since several decay channels might be opened with particle emission times usually shorter than ~10$^{-21}$ seconds. 
   Among these broad states, one might find narrow resonances \cite{oko1,PhysRevC.67.054322,bic2} which are the principal source of information about the spectroscopic properties and clusterization of unbound nuclei.
 
  The interplay between Hermitian and anti-Hermitian continuum couplings is a general mechanism of the formation of the narrow resonances in the continuum. Specifically, in the proximity of the decay threshold, the mixing of SM eigenstates leads to the formation of a collective eigenstate of the OQS Hamiltonian which carries an imprint of the decay channel \cite{okolowicz2012,okolowicz2013}. Numerous examples of such states are known in light nuclei. Among them are near-threshold $\alpha$-cluster states \cite{Ikeda1968}, such as the famous $0^+_2$ Hoyle state in $^{12}$C, neutron halo states, such as $^{11}$Li and $^{11}$Be, or the narrow resonance $1/2^-_1$ in $^{15}$F near the 2p-decay threshold \cite{de2016above}.

  Three unusually narrow resonances in the unbound $^{15}$F nucleus have been predicted by Canton et al. \cite{Canton06} and updated predictions are given in Ref.~\cite{PhysRevC.100.024609, Fortune07, Fortune11}. The prediction of these resonances was partially confirmed by the experimental observation of a narrow ($\Gamma$~=~36(19)~keV) resonance located only 129~keV above the 2p-decay threshold  \cite{de2016above}. 

  In the present work we report, for the first time, the clear observation of two new narrow resonances in $^{15}$F more than 3~MeV above the Coulomb barrier, by the resonant elastic ($^{1}$H($^{14}$O,p)$^{14}$O) and inelastic scattering ($^{1}$H($^{14}$O,2p)$^{13}$N) reactions. The spectroscopic properties of these resonances have been determined from a phenomenological R-matrix analysis of the excitation functions of these reactions.

  \begin{figure*}[ht]
    \begin{center}
      \includegraphics[width=\textwidth, angle=0]{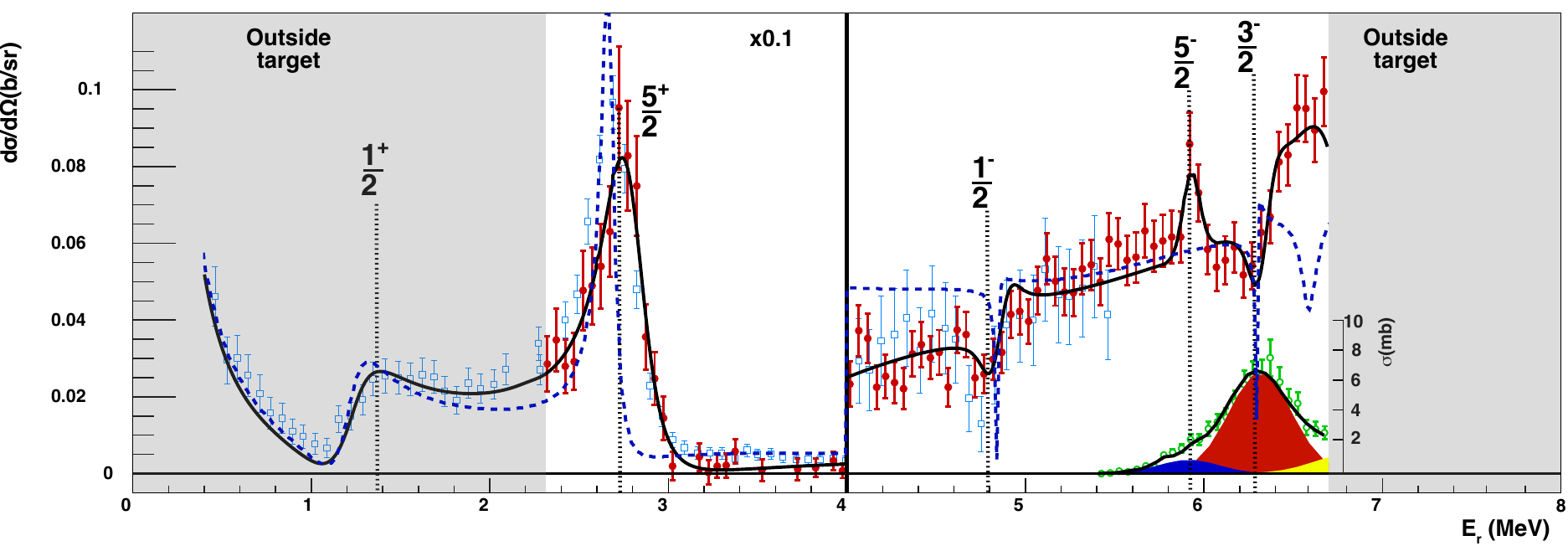} 
    \end{center}
    \vskip -0.5truecm
    \caption{(Color online) 
    Differential cross section of the $^{1}$H($^{14}$O,p)$^{14}$O reaction measured in the present study (full red dots) and in Ref. \cite{de2016above} (empty blue squares) and total cross-section of the $^{1}$H($^{14}$O,2p)$^{13}$N reaction (empty green circles), both, as a function of the reconstructed resonance energy $E_r$ in the $p$+$^{14}$O system. For the latter, the contribution of the 5/2$^{-}$, 3/2$^{-}$ states and higher-energy resonances extracted from the R-Matrix fit are shown in filled blue, red and green respectively. 
    The best R-matrix simultaneous fit of the two reaction channels constrained only by the g.s. properties extracted from Ref. \cite{de2016above} is shown as a continuous black line. The blue dashed line corresponds to the result of the GSMCC calculation (see text for details) scaled in amplitude to match the experimental cross-section.}
    \label{fig2}
    \vskip -0.5truecm
  \end{figure*}
  
  \textit{Experimental method.--} The experimental results have been obtained from a campaign of two measurements performed at GANIL using $^{14}$O radioactive beam delivered by the SPIRAL1 facility. 
  The unbound nucleus $^{15}$F was studied through the measurement of the  $^{1}$H($^{14}$O,p)$^{14}$O and $^{1}$H($^{14}$O,2p)$^{13}$N reactions. Both measurements used the thick-target technique \cite{de2016above}. The first measurement used a 7.64(1)~MeV/u beam of $^{14}$O impinging a 107(11)~$\mu$m-thick CH$_2$ target while the second experiment used a 7.42(1)~MeV/u $^{14}$O beam impinging on a 92(9)~$\mu$m-thick CH$_2$ target. A 75(8)~$\mu$m-thick $^{12}$C target was used to determine and subtract the carbon-induced background. A low-pressure multiwire detector, CATS \cite{CATS}, located upstream of the target was used in both experiments to monitor the beam intensity ($\sim$ 3$\times$10$^{5}$~pps). 

  \par
  The $^{1}$H($^{14}$O,p)$^{14}$O excitation function has been obtained in the first experiment \cite{E744Lise} from a MUST2 detector \cite{must2} responsible for the particle identification and the measurement of the total energy and angle of the protons. This telescope was composed of two stages: a square 300~$\mu$m-thick DSSD with 128x128 strips, and a 4x4 CsI crystals array and covered angles between 0$^{\circ}$ and 5$^{\circ}$ relative to the beam direction. A 57(5)~$\mu$m Ta foil acted as a beam stopper completely stopping beam-like particles from entering the detector, while having a minimal effect on the elastically scattered protons. 

  The $^{1}$H($^{14}$O,2p)$^{13}$N excitation function has been measured in the second experiment \cite{E744Vamos} using the invariant mass method. This experiment used for the first time the recently commissioned, state-of-the-art, detection system composed of the MUGAST array \cite{MUGAST} which includes 4 MUST2 detectors, the VAMOS magnetic spectrometer \cite{VAMOS} and the HPGe $\gamma$-ray spectrometer AGATA \cite{AGATA}. This detection system allowed the measurement of the full kinematics of the reaction. Particle identification, total energy and angle of the protons have been obtained using MUST2 telescopes covering angles between 8 and 50 degrees in the lab relative to beam direction. For beam-like residues, such as $^{13}$N, their total energy, angle and identification have been obtained using VAMOS with an acceptance up to 4.6 degrees relative to the beam direction. The $\gamma$-rays from the decay of unbound states were detected by AGATA but they are not discussed in this paper. 

  \par
  \textit{Experimental results.--} The measured excitation functions for $^{1}$H($^{14}$O,p)$^{14}$O  and  $^{1}$H($^{14}$O,2p)$^{13}$N reactions are shown in Fig. \ref{fig2} and the determined properties of the resonances are summarized in Table \ref{Results_1p2p}. The analysis of the two excitation functions has been performed using R-matrix formalism \cite{lane} implemented into the \emph{AZURE2} code \cite{azure} (radius parameter $a=5.1$~fm). The center-of-mass energy resolution considered for the resonant elastic and inelastic excitation function (see Fig.~\ref{fig2}) are respectively: $\sigma(E_{r})$~=~50(5)~keV and $\sigma(E_{r})$~=~300(20)~keV. 
  The experimental spectroscopic factors were deduced from the measured partial width $\Gamma(E)$ and the single particle width $\Gamma_{\rm sp}(E)$ (calculated with the \emph{DWU} code \cite{dwu}): $C^2S_{\rm exp}~=~\Gamma(E)/\Gamma_{\rm sp}(E)$. The experimental spectroscopic factors to the ground and first-excited states of $^{14}$O are displayed in Table \ref{table3}.   

  \begin{table}
    \caption{Resonances properties determined from the R-matrix analysis of the resonant elastic (${}^{15}$F$\rightarrow{}^{14}$O$({0^+_1})+{\rm p}$) and inelastic scattering (${}^{15}$F$\rightarrow{}^{14}$O$(1^-)+{\rm p}$) excitation functions. 
      }
    \begin{ruledtabular}
      \begin{tabular}{cc|cc|cc}
        $J^{\pi}$                   &  E$_{r}$(MeV)  & \multicolumn{2}{c|}{${}^{15}$F$\rightarrow{}^{14}$O${(0^+_1)}+{\rm p}$}     &    \multicolumn{2}{c}{${}^{15}$F$\rightarrow{}^{14}$O${(1^-_1)}+{\rm p}$} \\
        &                &    $\Gamma$ (keV)   & $\ell$  &     $\Gamma$ (keV)     & $\ell$  \\
        \hline
        5/2$^+$                     & 2.81(12)     & 251(26)    & 2   & -       & - \\ 
        1/2$^-$                     & 4.88(21)     & 30(15)     & 1  & -       &  - \\
        5/2$^-$                     & 5.93(10)     & 3(2)       & 3 & 0.3 (1) & 2 \\
        3/2$^-$                     & 6.33(13)     & 28(13)     & 1 & 2.2 (6) & 2 \\ 
        & &      &   & $<$ 1$\times$10$^{-3}$  &  0 \\ 
      \end{tabular}
    \end{ruledtabular}
    \label{Results_1p2p}
    \vskip -0.6truecm
  \end{table}

  The ground state (g.s.) is a broad resonance J$^{\pi}$~=~$1/2^+$ \cite{Fortune06,Mukhamedzhanov10,Baye05,Kekelis78,Benenson78,Peters03,Lepine03,Lepine04,Goldberg04,Guo05,de2016above,Fortune06}, closely related to the configuration [$^{14}$O${(0_1^+)}$ + ${\rm p}(s_{1/2})$]. 
  The first excited state (J$^{\pi}$~=~$5/2^+$,  E$_{r}$~=~2.81(12)~MeV, $\Gamma$~=~251(26)~keV) is in good agreement with previous measurements \cite{de2016above}. Based on the large spectroscopic factor C$^2$S~=~1.0 (see Table \ref{table3}), its structure is interpreted as
[$^{14}$O${(0_1^+)}$ + ${\rm p}(d_{5/2})$] \cite{Fortune05,Fortune06,Grigorenko15}.

  \par
  Contrary to the positive-parity resonances, the $1/2^-_1$, $5/2^-_1$, $3/2^-_1$ inherit weakly from the $^{14}$O${(0_1^+)} + {\rm p}$ configuration. Indeed, these states are collectivized by the coupling to 2p-decay channel $^{13}$N + 2p and to several inelastic 1p-decay channels. The second excited state J$^{\pi}=1/2^{-}$, has been found at E$_r$~=~4.88(21)~MeV, $\Gamma$~=~30(15)~keV, confirming previous measurements \cite{de2016above,Lepine04,Mukha09,Valerian21}. The small decay width of this resonance, which is situated more than 1.5~MeV above the Coulomb plus centrifugal barrier and almost 4.9~MeV above the 1p-emission threshold, has been explained \cite{de2016above} as a consequence of the continuum-coupling induced collective mixing of SM eigenstates  \cite{Oko2012,Oko2013} with the nearby 2p-decay channel.

  \begin{figure}
    \begin{center}
      \includegraphics[width=0.5\textwidth, angle=0]{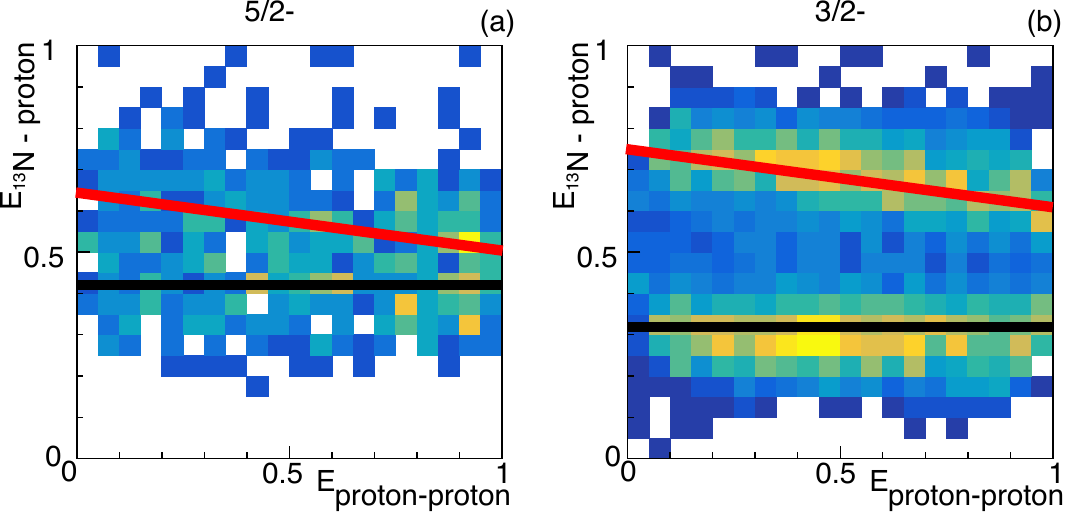} 
    \end{center}
    \vskip -0.6truecm
    \caption{(Color online) Reduced relative energy between one proton and the $^{13}$N residue is shown vs reduced relative energy between the two protons for the $^{1}$H($^{14}$O,2p)$^{13}$N events. The red and black lines corresponds respectively to the first and the second sequentially emitted protons.}
    \label{Dalitz15F}
    \vskip -0.6truecm
  \end{figure}

  \par
  At higher excitation energies, two new narrow resonances have been measured for the first time. These states can decay by one-proton emission to $^{14}$O${(0_1^+)}$ or sequentially by two-proton emission to the g.s. of $^{13}$N${(1/2^-_1)}$ through the intermediate state $^{14}$O$({1_1^{-}}$).  The 2p-decay has been investigated by the mean of the invariant-mass method. A comparison of the Dalitz plot representation \cite{perkins2000} of the experimental events compared to a realistic \emph{GEANT4} \cite{G4} simulation performed within the \emph{nptool} framework \cite{nptool} (see Fig. \ref{Dalitz15F}) indicated that the sequential 2p-decay through the first excited state of $^{14}$O dominates.

  \par
  The first new state is observed as a very narrow resonance with J$^{\pi}$ = $5/2^{-}$, E$_r$~=~5.93(10)~MeV, $\Gamma(0_1^+)$~=~3(2)~keV and $\Gamma(1^-_1)$~=~0.3(1)~keV. The obtained spectroscopic information for this resonance is consistent with the mirror-nucleus level sequence and the prediction of Fortune and Sherr \cite{Fortune07}. 

  \begin{figure}
    \begin{center}
      \includegraphics[width=0.5\textwidth, angle=0]{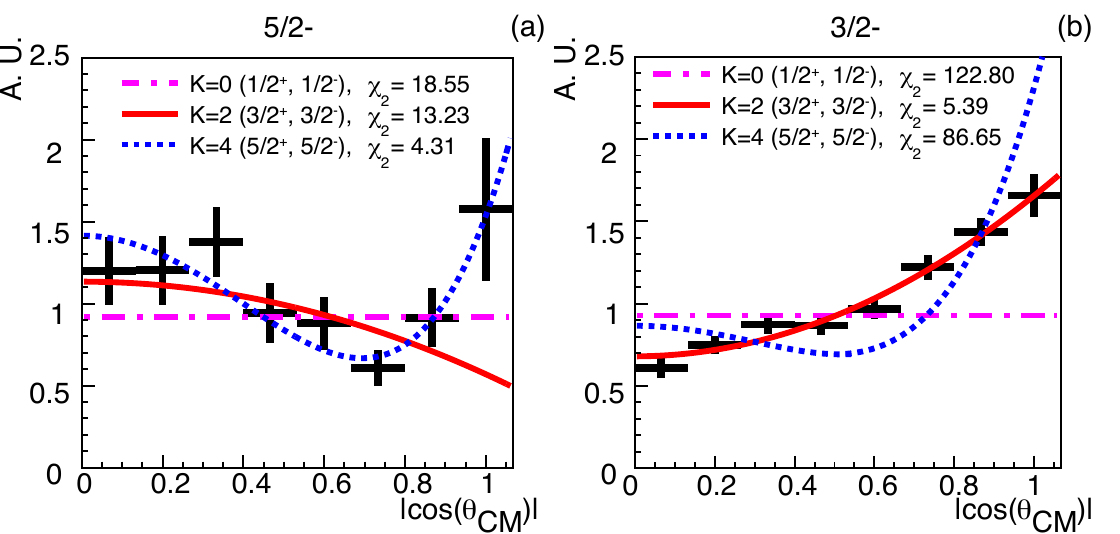} 
    \end{center}
    \vskip -0.6truecm
    \caption{(Color online) (a), (b): Measured center-of-mass angular distribution of the first sequentially emitted proton reaction (black dots). The red line, the dot-dashed pink line and the dotted blue line display the best Legendre polynomials fit with the order K=0, K=2 and K=4 \cite{Pronko}}
    \label{AngCorr}
    \vskip -0.6truecm
  \end{figure}

  \par
  The second new state is a 3/2$^-$ resonance with E$_r$~=~6.33(13)~MeV, $\Gamma(0_1^+)$~=~28(13)~keV and $\Gamma(^-_1)$ = 2.2(6)~keV. The R-matrix analysis could not distinguish between 3/2$^{-}$ and 1/2$^{+}$ spin-parity assignments for this resonance. The 3/2$^{-}$ spin assignment for this resonance is based on the Legendre polynomial fit \cite{Pronko} of the center-of-mass angular distribution of the first sequentially emitted proton in the $^{1}$H($^{14}$O,2p)$^{13}$N reaction as shown in Fig. \ref{AngCorr}. The analysis of this angular distribution also showed that the decay was dominated by the l=2 component with a $\ell$~=~0 partial width $<$ 1 eV (see Fig. \ref{AngCorr} and Table \ref{Results_1p2p}).
  The energy and spin-parity of this resonance agrees with a prediction of Ref. \cite{Fortune07, Fortune11}. However, the width is surprisingly 12.5 times narrower than the predicted value from Ref. \cite{Fortune11}, where they deduce the width from the $^{15}$C mirror nucleus  using the usual equality of spectroscopic factors between mirror states, without considering the coupling to the continuum. 

  \par
  \textit{Theoretical description.--}
  The present data has been analysed and interpreted in the framework of the Gamow Shell Model (GSM) \cite{Michel2002,PhysRevLett.89.042501,PhysRevC.67.054311,Michel2009,GSMbook}. This model is a configuration-interaction OQS approach formulated in the Berggren single-particle (s.p.) basis \cite{BERGGREN1968265} that includes bound states, resonances and non-resonant background states of the discretized contour embedding resonances. The many-body basis states consist of Slater determinants where nucleons occupy s.p. basis of the chosen Berggren basis \cite{Michel2009,GSMbook}. 

  \par
  The model space consists of $^{12}$C core and valence protons in the ${ p }$ and ${ sd }$ shells.  The s.p. valence space is build of three pole states: ${ 0p_{1/2}, 0d_{5/2} }$, ${ 1s_{1/2} }$ and five continua: ${ p_{1/2}, p_{3/2}, d_{3/2}, s_{1/2} }$, and ${ d_{5/2} }$. The $d_{5/2}$ and $s_{1/2}$ levels are described in the Berggren s.p. basis, whereas remaining partial waves are expanded in the harmonic oscillator basis. The Hamiltonian consists of a one-body part which includes a Woods-Saxon, spin-orbit and Coulomb potentials, and a two-body part which comprise of the Furutani-Horiuchi-Tamagaki effective interaction plus the Coulomb interaction and recoil term \cite{PhysRevC.91.034609}. Parameters of the Hamiltonian are adjusted to reproduce the energies of $1/2^-_1$,  $1/2^+_1$, $5/2^+_1$ states of $^{13}$N and $^{15}$F, as well as the energies of $0^+_1$, $1^-_1$, and $3^-_1$ states in $^{14}$O. 

  \par
  To describe nuclear reactions, GSM has to be formulated in the coupled-channel representation, the GSMCC \cite{PhysRevC.89.034624,PhysRevC.99.044606,GSMbook}.
  In the present studies, the reaction channels in the GSMCC are constructed by coupling states of ${ {  }^{ 14 } }$O calculated in the GSM with  proton states in different partial waves ${ (n \ell j) }$. The considered states of ${ { }^{ 14 } }$O are ${ { 0 }_{1}^{ + }}, { { 0 }_{2}^{ + }}, { { 2 }_{1}^{ + }}, { { 2 }_{2}^{ + }}, { { 0 }_{1}^{ - }}, { 1 }_{ 1 }^{ - }, { 2 }_{ 1 }^{ - }$, and ${ 3 }_{ 1 }^{ - } $. The projectile motion is described by single particle states of ${ p }$, and ${ sd }$ shells, and by ${ s }$, ${ p }$ and ${ sd }$ continua. 

  \begin{figure}
    \begin{center}
      \includegraphics[width=0.5\textwidth, angle=0]{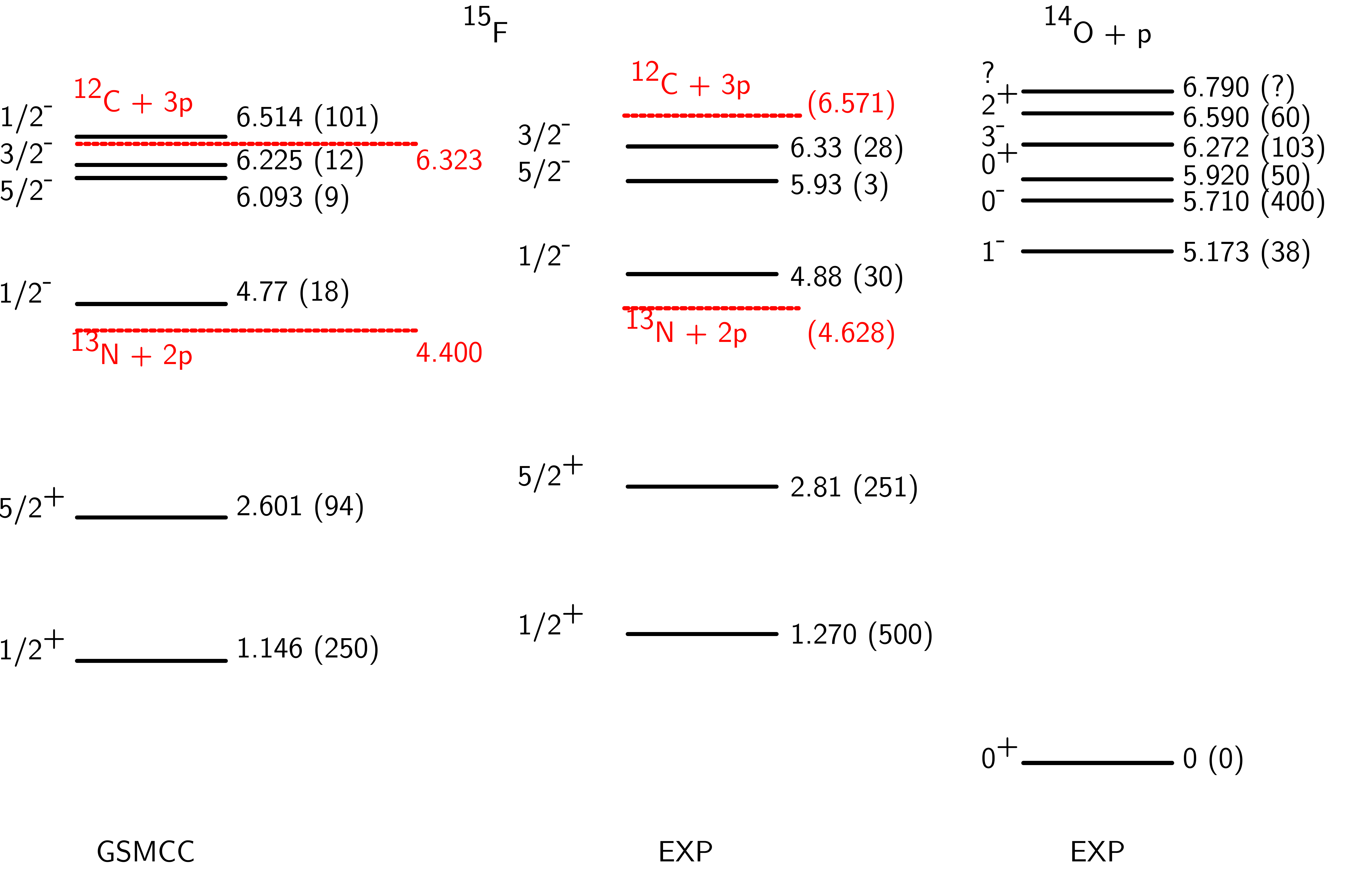}
    \end{center}
    \vskip -0.5truecm
    \caption{ (Color online) Level scheme of $^{15}$F with respect to $^{14}$O g.s. from the GSMCC calculation (left) and the present analysis (center), and level scheme of $^{14}$O (right). $^{15}$F data are taken from \cite{de2016above} and the present analysis, and $^{14}$O data are taken from \cite{NNDC}. The red lines indicate the 2p and 3p decay thresholds. Resonance widths are given in the brackets.  }
    \label{figure1}
    \vskip -0.5truecm
  \end{figure}

  \par
  The GSMCC excitation function is superimposed and scaled in amplitude to match the data in Fig. \ref{fig2}. Strong resonances are seen for the positive-parity states $J^{\pi} = 1/2^+_1, 5/2^+_1$ of $^{15}$F \cite{de2016above}. Excitation of the negative-parity states $1/2^-_1$ and $3/2^-_1$ is weak in accordance with both their small spectroscopic factors (see Table \ref{table3}) and the structure of their wave functions (see Table \ref{table1}) which do not resemble the wave function of a proton outside the $^{14}$O$(0_1^+)$ core. 

  \begin{table}
    \caption{\label{table_spectra} Major amplitudes of channels $\left[^{14}{\rm O}(K^{\pi})\otimes (\ell_{j})\right]^{J^{\pi}}$ in the lowest GSMCC resonances of $^{15}$F. 
    ${\cal R}$[${\tilde c}$] denotes real part of the channel amplitude. ${\cal R}$[${\rm S}$] corresponds to the 
    real part of the spectroscopic factor in open channels and C$^2$S$_{\rm exp}$ display the experimental spectroscopic factors, see text for details.} \label{table3}
    \begin{ruledtabular}
      \begin{tabular}{cccccc}
        $^{15}$F ; J$^{\pi}$ & $^{14}$O ; K$^{\pi}$ & $\ell_j$ & ${\cal R}$[${\tilde c}$] & ${\cal R}$[${\rm S}$]  & C$^2$S$_{\rm exp}$ \\
        \hline 
        ${ {1/2}_{1}^{+} }$ & ${ {0}_{1}^{+} }$ & ${ {s}_{1/2} }$ & 0.57 & 0.985   & 1  \\
        & ${ {1}_{1}^{-} }$ & ${ {p}_{1/2} }$ & 0.32  &  -   & -  \\
        & ${ {0}_{1}^{-} }$ & ${ {p}_{1/2} }$ & 0.1  &   -   & -   \\
        \hline
        ${ {5/2}_{1}^{+} }$ & ${ {0}_{1}^{+} }$ & ${ {d}_{5/2} }$ & 0.43  & 0.979 & 1   \\
        & ${ {3}_{1}^{-} }$ & ${ {p}_{1/2} }$ & 0.33  &  -  & -   \\
        & ${ {2}_{1}^{-} }$ & ${ {p}_{1/2} }$ & 0.23  &  -  & -    \\
        \hline
        ${ {1/2}_{1}^{-} }$ & ${ {0}_{1}^{+} }$ & ${ {p}_{1/2} }$ & 1.4$\times$10$^{-2}$  & 3.2$\times$10$^{-2}$ & 6$\times$10$^{-3}$  \\
        & ${ {0}_{2}^{+} }$ & ${ {p}_{1/2} }$ & 0.49  & - &  -     \\
        & ${ {1}_{1}^{-} }$ & ${ {s}_{1/2} }$ & 0.34  & -  & -     \\
        & ${ {0}_{1}^{-} }$ & ${ {s}_{1/2} }$ & 0.11   &  - & -     \\
        \hline
        ${ {5/2}_{1}^{-} }$ & ${ {0}_{1}^{+} }$ & ${ {f}_{5/2} }$ & -  & - &  3$\times$10$^{-3}$  \\
        & ${ {2}_{1}^{+} }$ & ${ {p}_{1/2} }$ & 0.39  & - &  -    \\
        & ${ {3}_{1}^{-} }$ & ${ {s}_{1/2} }$ & 0.32  & - &  -      \\
        & ${ {1}_{1}^{-} }$ & ${ {d}_{5/2} }$ & 0.14  & 0.477   & 0.7  \\
        & ${ {0}_{1}^{-} }$ & ${ {d}_{5/2} }$ & 0.1  & 0.356    & -   \\
        \hline
        ${ {3/2}_{1}^{-} }$ & ${ {0}_{1}^{+} }$ & ${ {p}_{3/2} }$ & 5$\times$10$^{-4}$ & 9$\times$10$^{-3}$ &  1$\times$10$^{-3}$  \\
        & ${ {2}_{1}^{+} }$ & ${ {p}_{1/2} }$ & 0.39  &  - & -    \\
        & ${ {2}_{1}^{-} }$ & ${ {s}_{1/2} }$ & 0.32  &  - & -   \\
        & ${ {1}_{1}^{-} }$ & ${ {d}_{5/2} }$ & 0.25  & 0.838 & 0.3  \\
        & & ${ {s}_{1/2} }$ & 1.9$\times$10$^{-5}$  & $<$ 3$\times$10$^{-6}$ & $<$ 1.3$\times$10$^{-5}$     \\
      \end{tabular}
    \end{ruledtabular}

  \end{table}	

  The $f$ shell is absent in the model space, therefore, there is no contribution of the $5/2^-$ partial wave in the elastic scattering reaction. For the same reason, the channel $\left[^{14}{\rm O}(0_1^{+})\otimes (f_{5/2})\right]^{5/2^{-}}$ is absent in Table \ref{table3}. The slow rise of the cross-section at energies higher than 6.5~MeV is due to higher energy resonances consisting of mainly a broad $3/2^+$ partial wave.

    \begin{table}
    \caption{\label{table_spectra} Number of protons in resonant shells: $\ell=0$ (s), $\ell=1$ (p), $\ell=2$ (d), and non-resonant shells of the scattering continuum: $\ell=0$ ($\{$s$\}$), $\ell=1$ ($\{$p$\}$), $\ell=2$ ($\{$d$\}$) for largest configurations in different eigenfunctions of $^{15}$F, $^{14}$O, and $^{13}$N. ${\cal R}$[c]  denote the real parts of the studied configuration amplitude.} \label{table1}
    \begin{ruledtabular}
      \begin{tabular}{ccccccccc}
        $^{15}$F  & J$^{\pi}$ & p & d & s & $\{$p$\}$ & $\{$d$\}$ & $\{$s$\}$ & ${\cal R}$[c]  \\ 
        \hline \\
        & 1/2${^+_1}$ & 2 & 0 & 1 & 0 & 0 & 0 & 0.89  \\
        & 5/2${^+_1}$ & 2 & 1 & 0 & 0 & 0 & 0 & 0.89  \\
        &  & 2 & 0 & 0 & 0 & 1 & 0 & 0.01  \\
        & 1/2${^-_1}$ & 1 & 0 & 2 & 0 & 0 & 0 & 0.82  \\
        &  & 1 & 2 & 0 & 0 & 0 & 0 & 0.13  \\
        & 1/2${^-_2}$ & 1 & 2 & 0 & 0 & 0 & 0 & 0.85  \\
        &  & 1 & 0 & 2 & 0 & 0 & 0 & 0.13  \\
        & 5/2${^-_1}$ & 1 & 1 & 1 & 0 & 0 & 0 & 0.86  \\
        &  & 1 & 2 & 0 & 0 & 0 & 0 & 0.08 \\
        & 3/2${^-_1}$ & 1 & 1 & 1 & 0 & 0 & 0 & 0.86  \\
        &  & 1 & 2 & 0 & 0 & 0 & 0 & 0.07  \\
        \hline \\
        $^{14}$O  
        & 0${^+_1}$ & 2 & 0 & 0 & 0 & 0 & 0 & 0.9  \\
        & 0${^-_1}$ & 1 & 0 & 1 & 0 & 0 & 0 & 0.86  \\
        &  & 1 & 0 & 0 & 0 & 0 & 1 & 0.12  \\
        & 1${^-_1}$ & 1 & 0 & 1 & 0 & 0 & 0 & 0.87  \\
        &  & 1 & 0 & 0 & 0 & 0 & 1 & 0.11  \\
        & 3${^-_1}$ & 1 & 1 & 0 & 0 & 0 & 0 & 0.95  \\
        &  & 1 & 0 & 0 & 0 & 1 & 0 & 0.03  \\
        & 0${^+_2}$ & 0 & 0 & 2 & 0 & 0 & 0 & 0.88  \\
        &  & 0 & 2 & 0 & 0 & 0 & 0 & 0.07  \\
        & 2${^+_1}$ & 0 & 1 & 1 & 0 & 0 & 0 & 0.85  \\
        &  & 0 & 2 & 0 & 0 & 0 & 0 & 0.1  \\
        & 2${^-_1}$ & 1 & 1 & 0 & 0 & 0 & 0 & 0.95  \\
        &  & 1 & 0 & 0 & 0 & 1 & 0 & 0.02  \\
        \hline \\
        $^{13}$N & 1/2${^-_1}$ & 1 & 0 & 0 & 0 & 0 & 0 & 0.98  \\		
      \end{tabular}
    \end{ruledtabular}
  \end{table}		

  Not observed in the present experiment, in the vicinity of 3p-decay threshold, the GSMCC predicts a second $1/2^-_2$ resonance at an energy close to the maximal accessible experimental energy (see Fig. \ref{fig2}). The structure of this resonance is dominated by the 3p-emission channel $\left[^{12}{\rm C}(0^+_1)\otimes (0p_{1/2}(0d_{5/2})^2)\right]^{1/2^-}$ and nearby 1p-emission channels $\left[^{14}{\rm O}(3^-_1)\otimes (0d_{5/2})\right]^{1/2^-}$ and $\left[^{14}{\rm O}(2^-_1)\otimes (0d_{5/2})\right]^{1/2^-}$. The major proton configurations are $[p^{1}d^{2}]$ and $[p^{1}s^{2}]$ (see Table \ref{table1}). Therefore, one expects an admixture of the direct 2p-emission in the decay of $1/2^-_2$ resonance.  
  The GSMCC also predicts several resonances at even higher energies which are clustered in the vicinity of 1p-emission channel $^{14}{\rm O}(2^+_1)$ + p and 2p-emission channels: $^{13}{\rm N}(1/2^+_1)$~+~2p, $^{13}{\rm N}(3/2^-_1)$~+~2p, and $^{13}{\rm N}(5/2^+_1)$~+~2p. 

   The calculated spectrum for $^{15}$F obtained from the GSMCC analysis is compared to the experimental one in Fig. \ref{figure1}. The obtained major channel amplitudes $\left[^{14}{\rm O}(K^{\pi})\otimes (\ell_{j})\right]^{J^{\pi}}$ for the resonances are presented in Table \ref{table3} together with the asymptotic normalization coefficient for major open channels in the resonances $1/2^+_1$, $5/2^+_1$, $5/2^-_1$, and $3/2^-_1$.
  Occupancies of s.p. shells for dominant GSM configurations in the considered states of $^{15}$F, $^{14}$O, and $^{13}$N are shown in Table \ref{table3}. It should be stressed that both energy and width of $5/2^-_1$ and $3/2^-_1$ resonances are well reproduced by the GSMCC (see~Fig.~\ref{figure1}).

  \par
  \textit{Structure of $^{15}$F negative parity states.--} As discussed in Ref. \cite{de2016above}, the $1/2^-_1$ resonance is narrow because the structure of this state is strongly affected by the coupling to a nearby 2p-decay channel. The dominant proton configurations are $[p^1s^2]$ and $[p^1d^2]$ whereas the major proton configuration in the g.s. of $^{14}$O is $[p^2]$ (see Table \ref{table1}). Consequently, the 1p-decay width is strongly reduced and the observation of the direct 2p-decay to the g.s. of $^{13}$N  is virtually impossible due to small available phase space for this decay. 
  The higher energy $5/2 ^ -_1$ and $3/2 ^ -_ 1$ resonances have a similar decay limitation as the $ 1/2 ^ -_1$  resonance. The dominant proton configurations in these resonances is $[p^{1}d^{1}s^{1}] $, i.e. their 1p-decay width is strongly reduced. 

 One may notice a similarity between the $5/2^-_1$, and $3/2^-_1$ resonances which are dominated by the same channels: the open channel $\left[^{14}{\rm O}(1^-_1)\otimes (0d_{5/2})\right]^{3/2^-,5/2^-}$ and the closed channel
  $\left[^{14}{\rm O}(2^+_1)\otimes (0p_{1/2})\right]^{3/2^-,5/2^-}$.  The principal difference  between these resonances is seen in the different $\ell = 0$ closed channels: $\left[^{14}{\rm O}(3^-_1)\otimes (1s_{1/2})\right]^{5/2^{-}_1}$ and $\left[^{14}{\rm O}(2^-_1)\otimes (1s_{1/2})\right]^{3/2^{-}_1}$, for states $5/2^-_1$ and $3/2^-_1$, respectively. 
  
 The couplings of the resonances $1/2^-_1$, $5/2^-_1$, and $3/2^-_1$ to the g.s. of $^{14}$O are extremely small and have been omitted from Table \ref{table3}. Small values of the experimental spectroscopic factors to the g.s. of $^{14}$O agree with the smallness of these couplings.
  One may also notice that the large GSMCC spectroscopic factor to the $1^-_1$ state of $^{14}$O for the $5/2^-_1$, and $3/2^-_1$ resonances, are in qualitative agreement with the obtained experimental values (see Table \ref{table3}).

  \par
  For the $1/2^-_1$ state, the $\left[^{14}{\rm O}(1^-_1)\otimes (0s_{1/2})\right]^{1/2^{-}}$ channel is closed and the sequential 2p-emission is not allowed. On the other hand, for $5/2^-_1$ and $3/2^-_1$ resonances the sequential 2p decay is possible via the intermediate $1^-_1$ state of $^{14}$O. This decay involves the major configurations of $5/2^-_1$, $3/2^-_1$ and $1^-_1$ states in $^{15}$F and $^{14}$O, respectively (see Fig. \ref{figure1} and Table \ref{table1}).

  \par
  \textit{Conclusions.--} The  negative parity states $1/2^-$, $5/2^-_1$, $3/2^-_1$ in the continuum of $^{15}$F 
  form a unique triplet of narrow resonances between 2p- and 3p-emission thresholds.
  Detailed GSM and GSMCC analysis provided an understanding of the imprint of the 2p-decay channel and various 1p-inelastic channels on the structure of those states.  
    The long lifetimes of $1/2^-$, $5/2^-_1$, $3/2^-_1$ resonances result from the vicinity of the decay channels: 
    $^{13}$N$(1/2^-_1)$ + 2p  and $^{14}$O${(1^-_1)}$ + p which collectivise these resonances and change their structure, thus preventing that the decay goes via the elastic channel $^{14}$O$(0^+_1)$ + p. For near-threshold resonances, the inference about their structure from the SM calculation of  (bound) mirror partners  becomes questionable. Near-threshold collectivization may play an important role in many reactions of astrophysical interest by modifying their rates with respect to the predictions based on the SM and the assumption of mirror symmetry \cite{Oliveira97}.
    
 \par
  The sequential 2p-decay of $5/2^-_1$ and $3/2^-_1$ states via $1^-_1$ resonance in $^{14}$O, has been observed, in agreement with the GSMCC calculations.  The direct 2p-decay component is predicted by the GSM to be present in the deexcitation of $1/2^-_2$ state just above the newly measured resonances. 

  \par
  Based on general theoretical arguments, one expects that narrow near-threshold resonances exist also in other nuclei beyond the proton and neutron drip lines. For example, the same sequence of narrow negative-parity resonances is expected in the unbound $^{13}$F and $^{17}$Na nuclei. Systematic investigations of the narrow resonances and their various particle and $\gamma$ decays modes in nuclei beyond drip lines will open new perspectives in the studies of effective interactions in nuclear OQS  \cite{GSMbook} and the continuum-coupling induced collectivization of the near-threshold states \cite{okolowicz2012,okolowicz2013}.

  \par
  \textit{Acknowledgements--}
  We thank the GANIL accelerator staff for delivering the radioactive beams. Special thanks go to the MUGAST, VAMOS and AGATA collaborations for their dedication. This work has been supported in part by the European Community FP6 Structuring the ERA Integrated Infrastructure Initiative, contract EURONS RII3-CT-2004-506065, U.S. Department of Energy grant No. DE-SC0019521, COPIN and COPIGAL French-Polish scientific exchange programs, the French-Romanian IN2P3-IFIN-HH collaboration No. 03-33, the French-Czech LEA NuAG collaboration, French-Serbian collaboration No. 20505 and MESTD of Serbia project No.171018, the polish grant NCN 2016/22/M/ST2/00269 and by the Helmholtz Association through the Nuclear Astrophysics Virtual Institute. This work has also benefited from high performance computational resources provided by Louisiana State University (www.hpc.lsu.edu).

   \bibliography{15F}

\end{document}